\documentclass[aps,prd,preprint,nofootinbib]{revtex4}
\usepackage{mathrsfs}
\usepackage{amsfonts}
\usepackage{epsfig}
\usepackage{graphicx}
\begin{document}
\title{The QCD NLO Corrections to Inclusive $B^*_c$ Production\\ in
$Z^{0}$ Decays\\[9mm]}

\author{\vspace{1cm} Jun Jiang$^1$\footnote{
jiangjun13b@mails.ucas.ac.cn}, Long-Bin Chen$^1$\footnote{
chenglogbin10@mails.ucas.ac.cn} and Cong-Feng
Qiao$^{1,2}$\footnote{qiaocf@ucas.ac.cn, corresponding author} \\}

\affiliation{$^1$School of Physics, University of Chinese Academy of Sciences, YuQuan Road 19A, Beijing 100049, China\\
$^2$CAS Center for Excellence in Particle Physics, Beijing 100049, China}


\begin{abstract}
\vspace{3mm} We calculate the next-to-leading order(NLO) quantum chromodynamics(QCD) corrections to the inclusive process of $Z_0 \rightarrow B^*_c+\bar{c}+b$ under the non-relativistic QCD(NRQCD) factorization scheme. Technical details about contributions from vector and axial-vector currents in dimensional regularization scheme are discussed. Numerical calculation shows that the NLO correction enhances the leading-order decay width by about 50\%, and the dependence on renormalization scale $\mu$ is reduced. The uncertainties induced by quark masses and the renormalization scale $\mu$ are also analyzed.

\vspace {7mm} \noindent {\bf PACS number(s):} 13.85.Ni, 14.40.Nd,
12.39.Jh, 12.38.Bx.

\end{abstract}

\maketitle
\section{Introduction}

Due to its unique nature in the family of mesons, $B_{c}$ system attracts wide attention of experiment and theory. The study of $B_{c}$ mesons may deepen our understanding of the Standard Model(SM) and the effective theory, non-relativistic quantum chromodynamics(NRQCD) \cite{NRQCD}. $B_{c}$ meson was first discovered at the Fermilab TEVATRON \cite{CDF}, and its excited state $B_c(2S)$ \cite{ATLAS} was recently observed by ATLAS Collaboration.

For $B_c$ meson direct production, various investigations had been carried out. Refs. \cite{Had,pwave,octet,intrinsic} studied in detail the $B_c$ hardroproduction, in which the P-wave \cite{pwave} and color-octet \cite{octet} contributions, as well as the ``intrinsic heavy quark mechanism" \cite{intrinsic} were taken into account. Refs. \cite{frag1,frag2} discussed the $B_c$ production through the fragmentation scheme, and a Monte-Carlo simulation program \cite{BCVEGPY} for the $B_c$ hadroproduction was also produced.

Apart from the direct production, $B_c$ meson indirect production is also interesting, which may inform us not only the nature of $B_c$ meson, but also the characters of its parent particles. $B_c$ production through the top quark decays was discussed in Refs. \cite{Top1,Top2}, through $W^{\pm}$ decays was calculated in Ref. \cite{W}, and through $Z^{0}$ decays was analyzed in Refs. \cite{frag2,Z1,Z2}. In the work of Ref. \cite{slp}, the NLO QCD corrections to $Z^{0}\rightarrow {B}_c(^1S_0)+\bar{c}+b$ process was calculated.

At the LHC and International Linear Collider(ILC) \cite{ILC}, or other forms of $Z$ factories, the $Z^0$ boson is and will be copiously produced.
The $Z^0$ production cross section at the LHC is about $34\text{nb}$ \cite{lhcz}, and at the ILC, e.g., the cross section will be about $30\text{nb}$ while collider runs at the $Z^0$ pole energy \cite{GigaZ}.
Given the colliders' luminosity to be $10^{34}\text{cm}^{-2}s^{-1} \approx10^8 \text{nb}^{-1}/\text{year}$, there will be $\sim10^{9}$ $Z^{0}$ events being produced per year at the LHC and ILC. Therefore, to study the $B_c$ production in $Z^0$ decays is worthwhile and meaningful. And, for this aim, since the $B_c^*$ will almost completely decay to scalar $B_c$ and the NLO QCD correction in heavy quarkonium energy region is large, in some cases even huge, in this work we calculate the NLO QCD corrections for $Z^{0}\rightarrow {B}_c^*(^3S_1)+\bar{c}+b$ process.

The rest of the paper is organized as follows. In section II we recalculate the $Z^0$ to $B^*_c$ process at the Born level. In section III, the NLO virtual and real QCD corrections to the leading order result are evaluated. Section IV presents some technical details of the calculation. In section V, the numerical evaluation for the concerned decay process is performed at NLO accuracy. The last section is for summary and conclusions.

\section{Leading Order Contribution}

\begin{figure}[t,m,u]
\centering
\includegraphics[width=16cm,height=3cm]{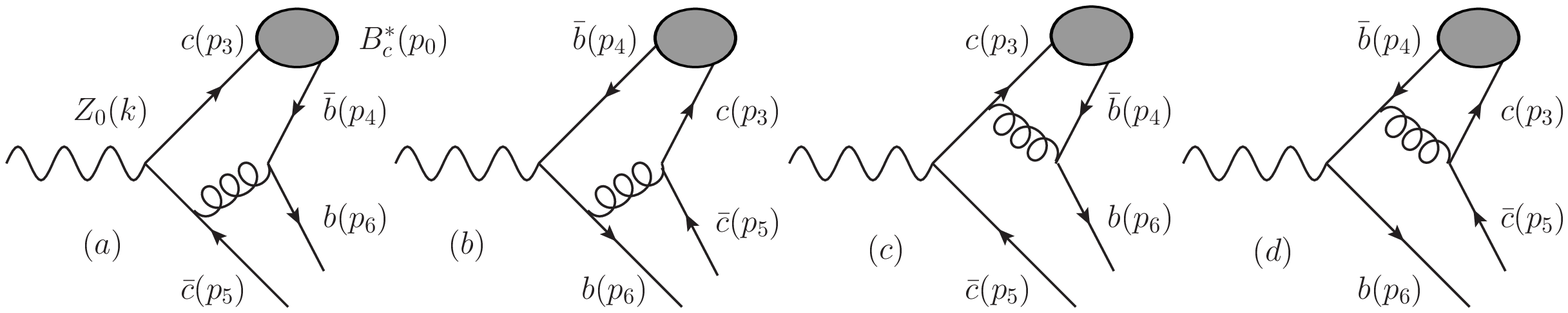}
\caption{\small The leading order Feynman diagrams for ${B}^*_c$
production in $Z^{0}$ decays.} \label{graph1}
\end{figure}

The $Z^{0}(k)\rightarrow {B}^*_c(p_{0})+\bar{c}(p_{5})+b(p_{6})$ process starts from $\alpha_{s}^2$ order, the Born level, and the corresponding four independent Feynman diagrams are shown in Fig.\ref{graph1}. The decay width may be expressed in a standard form, i.e.
\begin{eqnarray}
\mathrm{d}\Gamma_{Born}=\frac{1}{2m_{Z}}\frac{1}{3} \sum|{\cal
M}_{Born}|^{2}\mathrm{d}\textmd{PS}_{3}\ . \label{eqb:1}
\end{eqnarray}
Here, $\sum$ means summing over the polarizations and colors of
final particles, $1/3$ comes from the spin average
of initial state, and $\mathrm{d}\textmd{PS}_{3}$ stands for the
three-body phase space of final states, which can be explicitly expressed as
\begin{eqnarray}
\mathrm{d}\textmd{PS}_{3}=\frac{1}{2^7 \pi^3 m_Z^2}
\mathrm{d}s_{2}\mathrm{d}s_{1}\ , \label{eqb:2}
\end{eqnarray}
where $s_{1}=(p_{0}+p_{5})^{2}=(k-p_{6})^{2}$ and $s_{2}=
(p_{5}+p_{6})^{2}=(k-p_{0})^{2}$. The upper and lower bounds for $s_1$ and $s_2$ can be found in our previous work \cite{slp}, or any standard text book of particle physics.

The amplitude $M_{Born} = M_a + M_b + M_c + M_d$ can be readily obtained  from Fig.\ref{graph1} according to Feynman rules:
\begin{eqnarray}
{\cal M}_{a}&=&{\cal C}\times\bar{u}(p_{6})\gamma^{\mu}\frac{v(p_4)\bar{u}(p_3)}{(p_4+p_6)^2} {\not\!\epsilon(k)}(T_c -\gamma^5)\frac{(-\not\!p_4
-\not\!p_5-\not\!p_6)+m_c}{(-p_4-p_5-p_6)^2-m_c^2}\gamma_{\mu}v(p_{5})\ , \nonumber \\
\nonumber \\
{\cal M}_{b}&=&{\cal C}\times\bar{u}(p_{6})\gamma^{\mu}\frac{(\not\!p_3+\not\!p_5+\not\!p_6)+m_b}
{(p_3+p_5+p_6)^2-m_b^2}{\not\!\epsilon(k)}(\gamma^5-T_b) \frac{v(p_4)\bar{u}(p_3)}{(p_3+p_5)^2} \gamma_{\mu}v(p_{5})\ , \nonumber \\
\nonumber \\
{\cal M}_{c}&=&{\cal C}\times\bar{u}(p_{6})\gamma^{\mu}\frac{v(p_4)\bar{u}(p_3)}{(p_4+p_6)^2} \gamma_{\mu}\frac{(\not\!p_3+
\not\!p_4+\not\!p_6)+m_c}{(p_3+p_4+p_6)^2-m_c^2}{\not\!\epsilon(k)}(T_c -\gamma^5)v(p_{5})\ , \nonumber \\
\nonumber \\
{\cal M}_{d}&=&{\cal C}\times\bar{u}(p_{6}){\not\!\epsilon(k)} (\gamma^5-T_b)\frac{(-\not\!p_3-\not\!p_4-\not\!p_5)+m_b}
{(-p_3-p_4-p_5)^2-m_b^2}\gamma^{\mu}\frac{v(p_4)\bar{u}(p_3)}{(p_3+p_5)^2} \gamma_{\mu}v(p_{5})\ . \label{eqb:3}
\end{eqnarray}
Here, the constant ${\cal C}=\frac{\pi \alpha_s g C_F}{cos\theta_W}$ with $\theta_W$ being the Weinberg angle, $\epsilon(k)$ is the polarization vector of $Z^0$ boson, $T_c=(1 - \frac{8}{3}\sin^2\theta_W)$ and $T_b=(1 - \frac{4}{3}\sin^2\theta_W)$. For $c$ and $\bar{b}$ constituent quarks hadronization to $B_c^*$ meson, the following projection operator is employed \cite{ap}
\begin{eqnarray}
v(p_4)\bar{u}(p_3) \to \frac{\Psi_{^{3}S_{1}}(0)}{2\sqrt{m_{B^*_c}}}{\not\!\epsilon^*(p_0)}({\not\!p_0}
+m_{B_c^*}) \otimes \left( {{\bf
1}_c\over \sqrt{N_c}}\right)\ , \label{eqb:4}
\end{eqnarray}
where $\epsilon^*(p_0)$ is the polarization vector of $B^*_c$ meson with $p_0=p_3+p_4$, $m_{B_c^*}=m_c+m_b$, ${\bf 1}_c$ stands for the unit color matrix, and $\Psi_{^{3}S_{1}}(0)$, a nonperturbative parameter, is the Schr\"{o}dinger wave function at the origin of the $B_c^*$ meson.

\section{Next-to-Leading Order Corrections}

\begin{figure}[t,m,u]
\centering
\includegraphics[width=15cm,height=5cm]{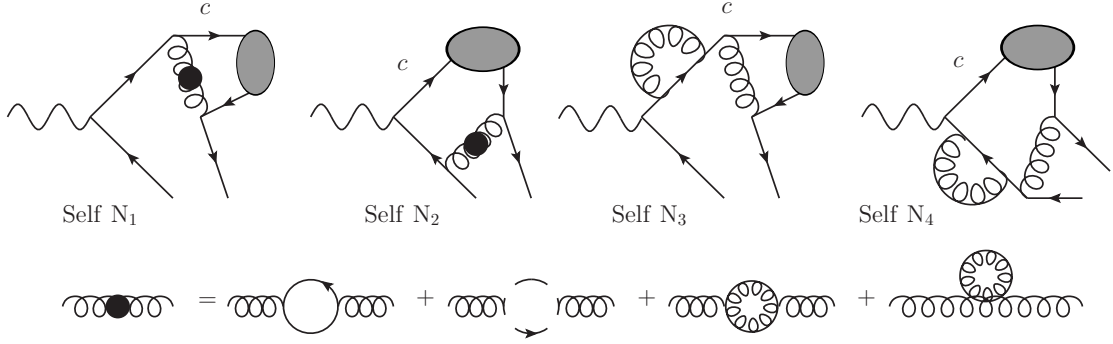}
\caption{\small Half of the self-energy diagrams in the virtual correction.}
\label{graph3}
\end{figure}

The NLO QCD corrections to the $Z^{0}(k)\rightarrow {B}^*_c(p_{0})+\bar{c}(p_{5})+b(p_{6})$ process contain virtual and real corrections, i.e. $\Gamma_{Virtual}$ and $\Gamma_{Real}$ respectively, which are both at the order of $\alpha_s^3$. Typical Feynman diagrams which attribute to the virtual correction are presented in Figs.\ref{graph2}- \ref{graph5}, while those for the real correction are shown in Fig.\ref{graph6}. Note that in Figs.\ref{graph3}- \ref{graph6} only those diagrams with $Z\rightarrow \bar{c}c$ vertex have been displayed, the remaining half can easily be obtained by exchanging the $c$ and $b$ quark lines.
\subsection{The Virtual Correction}

With virtual correction, the decay width can be formulated as
\begin{eqnarray}
\mathrm{d}\Gamma_{Virtual}=\frac{1}{2m_{Z}}\frac{1}{3}
\sum2\textmd{Re} ({\cal M}_{Virtual} {\cal M}_{Born}^{*})
\mathrm{d}\textmd{PS}_{3}\ .\label{eqv:1}
\end{eqnarray}
In $\textmd{Re} ({\cal M}_{Virtual} {\cal M}_{Born}^{*})$, both  ultraviolet(UV) and infrared(IR) divergences exist. The conventional dimensional regularization scheme wiht $D=4-2\epsilon$ is adopted to regularize them. There are also Coulomb divergences, which in this work are factorized out through the threshold expansion technique \cite{asymp} and then attributed to the bound state wave function. In the calculation,
the NRQCD short distance coefficients are obtained by matching to the full QCD result stemming solely from the hard interaction region, and other regions give no contribution.
\begin{figure}[t,m,u]
\centering
\includegraphics[width=16cm,height=2.5cm]{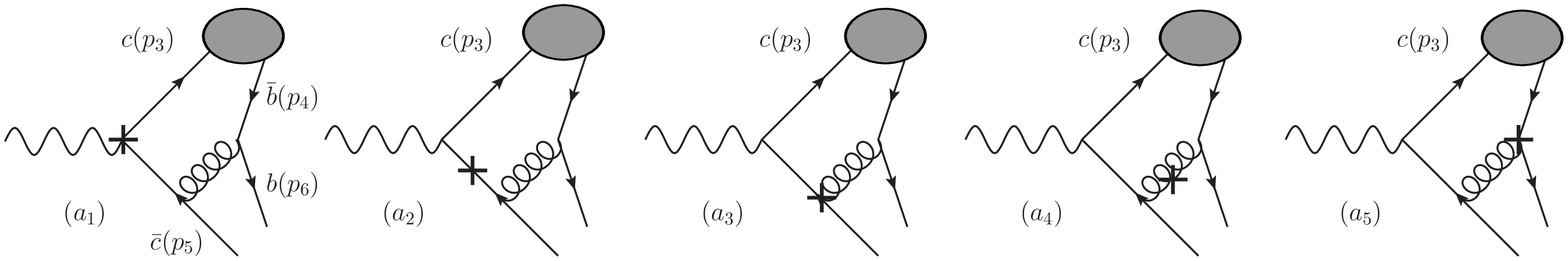}
\caption{\small The typical counter-term Feynman diagrams corresponding to Fig.\ref{graph1}$(a)$.} \label{graph2}
\end{figure}

According to the power counting rule, the UV divergences exist merely in self-energy and triangle diagrams, which are canceled by counter terms(CT). The Fig.\ref{graph2} contains 5 typical CT diagrams corresponding to the Fig.\ref{graph1}$(a)$, and hence there should be other 15 CT diagrams not shown. Of the 20 CTs, the renormalization constants include $Z_{2}$, $Z_{3}$, $Z_{m}$, and $Z_{g}$, corresponding to the renormalizations of quark field, gluon field, quark mass, and strong coupling constant $\alpha_{s}$, respectively. In practice, the terms related to $Z_{3}$ vanish, e.g. those contribute to $Z_{3}$ in Fig.\ref{graph2}, the $(a_3)$, $ (a_4)$ and $(a_5)$, cancel with each other. In the
calculation, $Z_{g}$ is defined in the modified-minimal-subtraction($\mathrm{\overline{MS}}$) scheme,
while for $Z_{2}$ and $Z_{m}$ we take the on-shell($\mathrm{OS}$). In the end of the day we have
\begin{eqnarray}
&&\hspace{-0.3cm}\delta Z_2^{OS}=-C_F\frac{\alpha_s}{4\pi}
\left[\frac{1}{\epsilon^{'}_{UV}} + \frac{2}{\epsilon^{'}_{IR}}-3\ln(m^2)+4\right]+{\mathcal{O}}(\alpha_s^2)\ , \nonumber\\
&&\hspace{-0.3cm}\delta Z_m^{OS}=-3C_F\frac{\alpha_s}{4\pi}
\left[\frac{1}{\epsilon^{'}_{UV}}-\ln(m^{2})+\frac{4}{3}\right]+ {\mathcal{O}}(\alpha_s^2)\ ,\nonumber\\
&&\hspace{-0.3cm}\delta Z_g^{\overline{MS}}=-\frac{\beta_0}{2}
\frac{\alpha_s}{4\pi}\left[\frac{1}{\epsilon^{'}_{UV}}-
\ln(\mu^2)\right]+{\mathcal{O}}(\alpha_s^2)\ .\label{eqv:2}
\end{eqnarray}
Here, $1/\epsilon^{'}_{UV(IR)} = 1/\epsilon - \gamma_E + \ln(4\pi\mu^2)$, $\mu$ is the renormalization scale, the mass $m$ in $\delta Z_2^{OS}$ and $\delta Z_m^{OS}$ stands for $m_{c}$ and $m_{b}$ accordingly, $\beta_{0}=(11/3)C_{A}-(4/3)T_{f}n_{f}$ is the one-loop coefficient of the QCD beta function with $n_{f} = 5$, the number of active quarks in our calculation. Eventually, the UV divergences appearing in CTs eliminate all UV divergences in self-energy and triangle diagrams, leading to a UV-free result. The remaining IR divergences in CTs have the following form
\begin{eqnarray}
\mathrm{d}\Gamma_{CT}^{IR}=-\frac{2C_F\alpha_s} {\pi\epsilon^{'}_{IR}}\mathrm{d}\Gamma_{Born}\ . \label{eqv:3}
\end{eqnarray}
\begin{figure}[t,m,u]
\centering
\includegraphics[width=15cm,height=9cm]{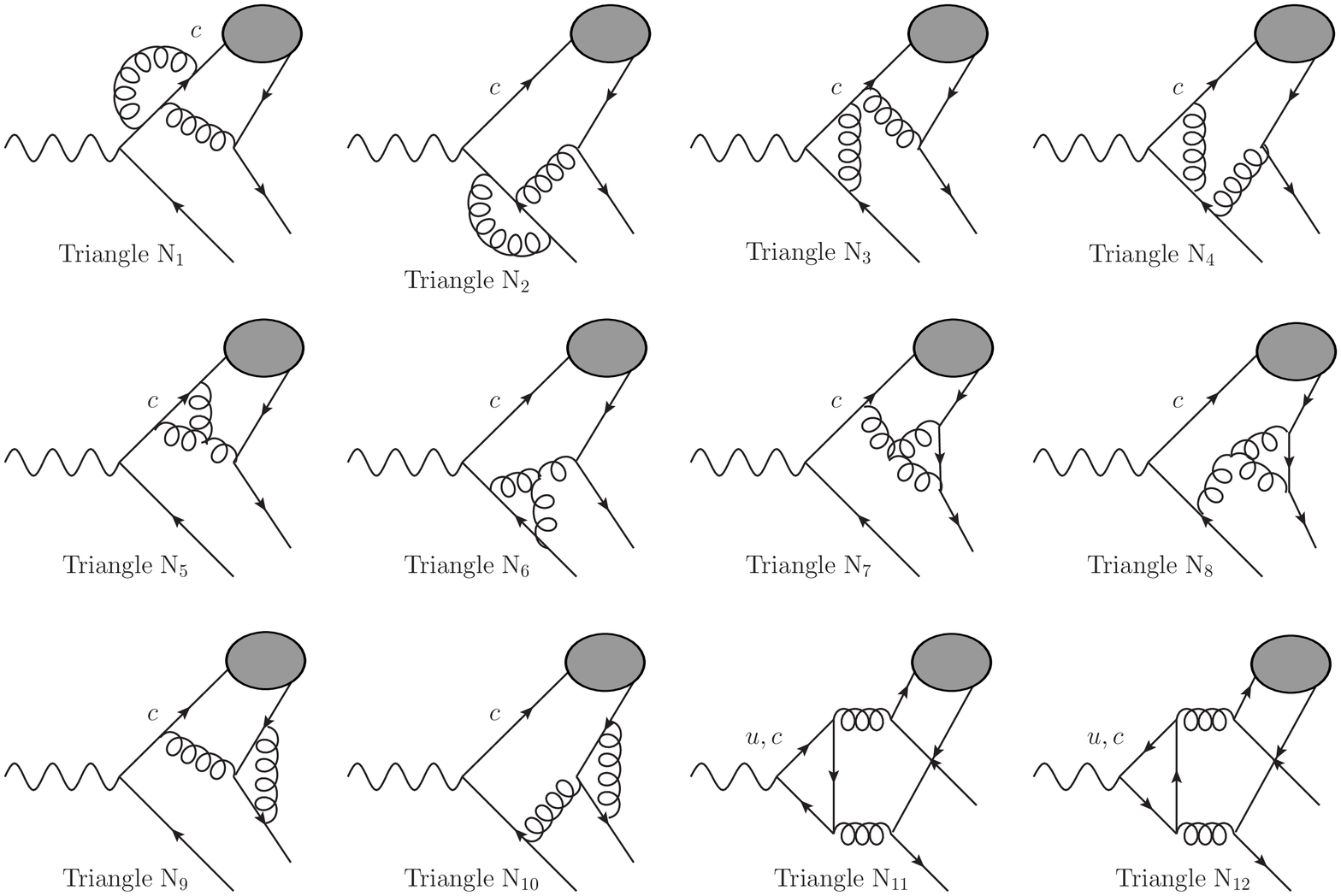}
\caption{\small Half of the triangle diagrams in the virtual correction.}
\label{graph4}
\end{figure}
\begin{figure}[t,m,u]
\centering
\includegraphics[width=15cm,height=11cm]{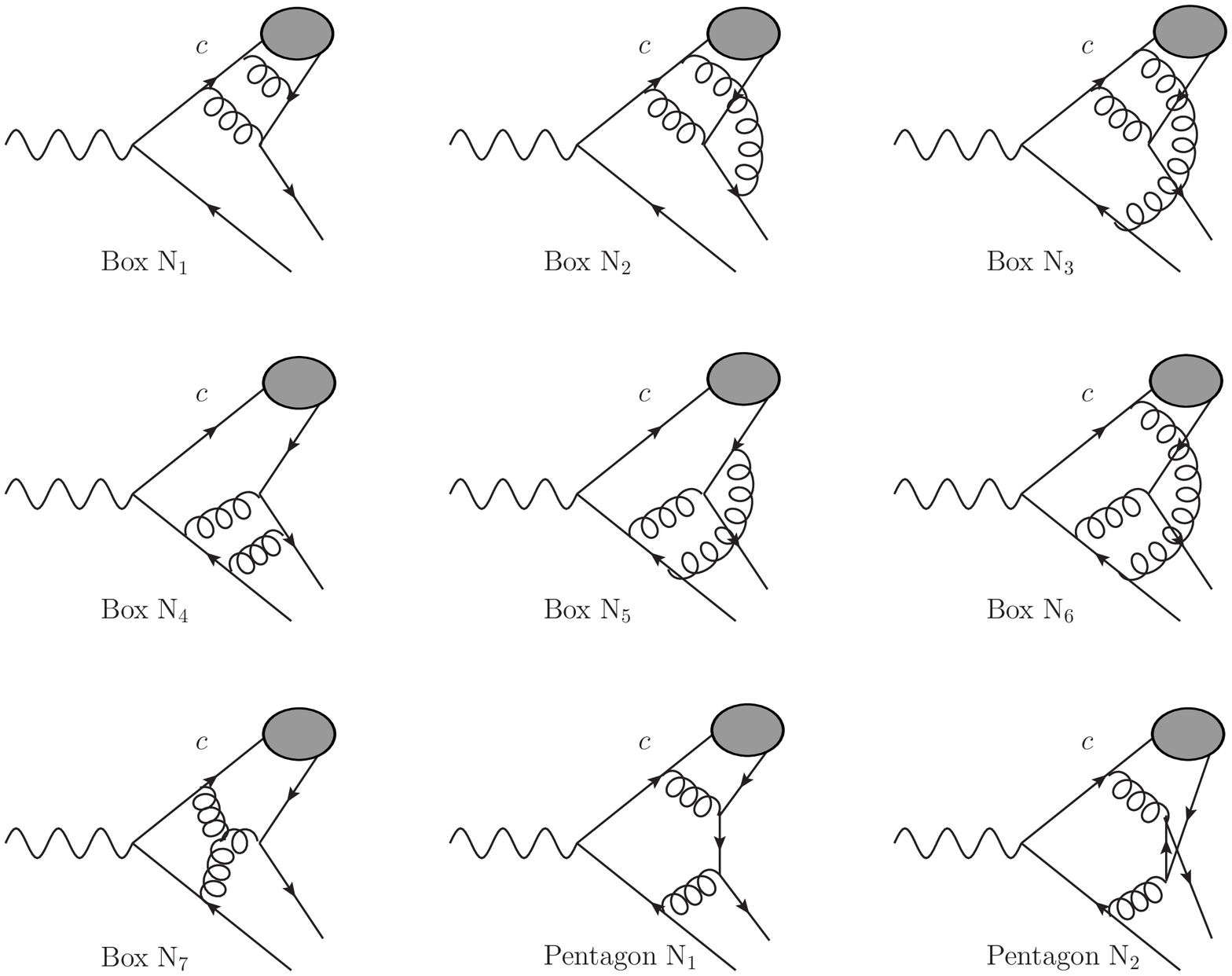}
\caption{\small Half of the box and pentagon diagrams in the virtual correction.} \label{graph5}
\end{figure}
In one-loop Feynman diagrams, the IR divergences involve triangle, box and pentagon diagrams. Of the triangle diagrams in Fig.\ref{graph4}, only two have IR divergences, i.e. $\mathrm{Triangle N_9}$ and $\mathrm{Triangle N_{10}}$. Of the diagrams in Fig.\ref{graph5}, $\mathrm{Box N_7}$ has no IR divergences, $\mathrm{Box N_1}$ and $\mathrm{Pentagon N_1}$ have both Coulomb singularities and IR divergences, and the remaining other diagrams have only the IR divergences. We find that the cancelation of IR divergences in Figs.(\ref{graph4}) and (\ref{graph5}) goes as follows:
\begin{itemize}
\item Combinations of $\mathrm{Triangle N_9+Box N_2}$, $\mathrm{Box N_5+Box N_6}$ are IR finite;
\item Combination of $\mathrm{Triangle N_{10}+Box N_3+Pentagon N_2}$ are IR finite;
\item The remaining IR divergences lie in $\mathrm{Box N_{1}}$, $\mathrm{Box N_4}$ and $\mathrm{Pentagon N_1}$;
\item Diagrams with $Z\rightarrow \bar{b}b$ vertex have the same cancelation pattern as in above.
\end{itemize}

\subsection{The Real Correction}

\begin{figure}[t,m,u]
\centering
\includegraphics[width=15cm,height=10cm]{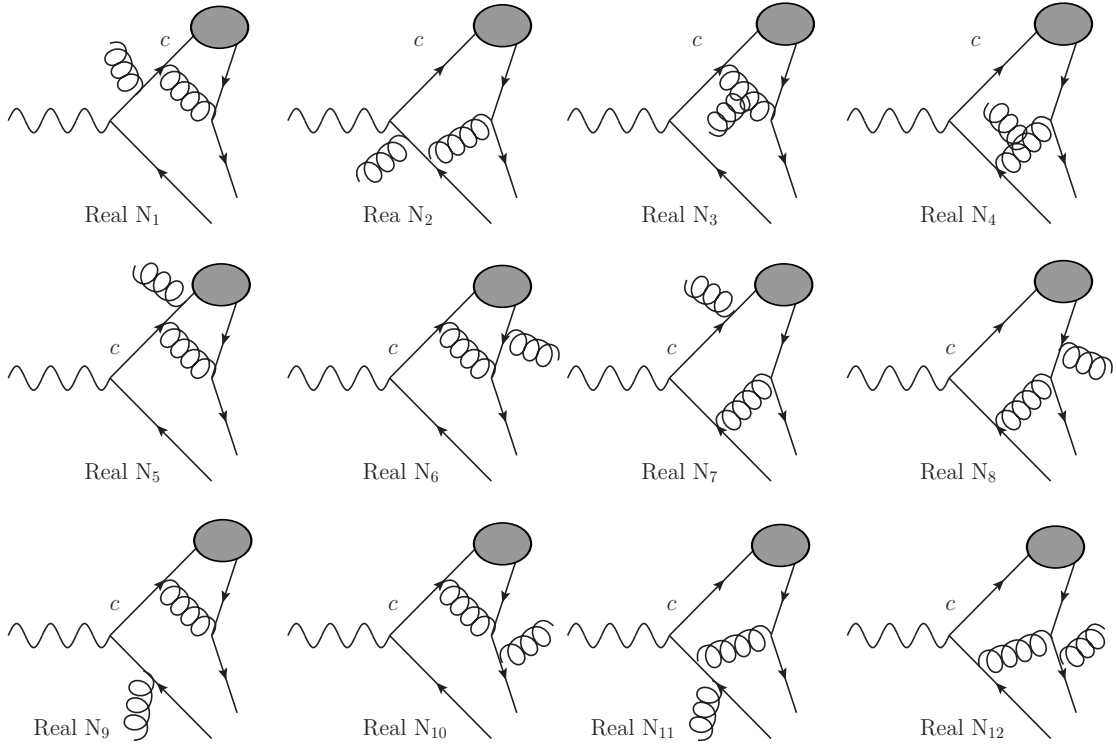}
\caption{\small Half of the Feynman diagrams in the real correction.} \label{graph6}
\end{figure}

For the real correction of the concerned process, in which the IR divergences exist, there are $24$ Feynman diagrams and half of them are shown in Fig.\ref{graph6}. To regularize the IR singularities, the ``Two cutoff phase space slicing method'' has been employed \cite{twocut}. For diagrams contain IR divergences, we enforce a cut $\delta$ on the
energy of the radiational gluon, the $p^0_{7}$. The gluon with energy $p_{7}^{0} < \delta$ is considered to be soft, while $p_{7}^{0} > \delta$ is treated as hard. Then, the decay width can be written as:
\begin{eqnarray}
&&\mathrm{d}\Gamma_{Real}=\mathrm{d}\Gamma_{Real}\mid_{IR}+\mathrm{d} \Gamma_{Real}\mid_{IR-free}\ ,\;\nonumber\\
&&\mathrm{d}\Gamma_{Real}\mid_{IR}=\mathrm{d} \Gamma^{IR-soft}_{Real}\mid_{p_{7}^{0} < \delta}+\mathrm{d}\Gamma^{IR-hard}_{Real}\mid_{p_{7}^{0} > \delta}\ .\label{eqr:1}
\end{eqnarray}
~~~~In Fig.\ref{graph6}, we find that
\begin{itemize}
\item The first 4 diagrams, the $\mathrm{Real N_{1,2,3,4}}$, are IR free;
\item The combination of $\mathrm{Real N_5+ Real N_6}$, $\mathrm{Real N_7+Real N_8}$ has no IR singularity;
\item The remaining 4 diagrams, i.e. $\mathrm{Real N_{9,10,11,12}}$, have IR divergences;
\item The diagrams with $Z\rightarrow \bar{b}b$ vertex behave the same as in above.
\end{itemize}
Then the decay width in the soft sector $\mathrm{d}\Gamma^{IR-soft}_{Real}\mid_{p_{7}^{0} < \delta}$ can be written as
\begin{eqnarray}
\mathrm{d}\Gamma_{Real}^{IR-soft}\mid_{p_{7}^{0} < \delta}=\frac{1}{2m_{Z}}\frac{1}{3} \sum|{\cal
M}^{IR-soft}_{Real}|^{2}\times\mathrm{d}\textmd{PS}_{4}\mid_{soft}. \label{eqr:2}
\end{eqnarray}
In the Eikonal approximation, the squared amplitude $|{\cal M}^{IR-soft}_{Real}|^{2}$ reads
\begin{eqnarray}
|{\cal M}^{IR-soft}_{Real}|^{2}=-(C_F 4\pi\alpha_s)|{\cal M}_{Born}|^{2}\times\left(\frac{p_5^2}{(p_5 \cdot p_7)^2}-2\frac{p_5 \cdot p_6}{(p_5 \cdot p_7)(p_6 \cdot p_7)}+\frac{p_6^2}{(p_6 \cdot p_7)^2}\right), \label{eqr:3}
\end{eqnarray}
which contains all the IR singularities in the real correction, i.e. $\mathrm{Real N_{9,10,11,12}}$ and the other 4 diagrams with $Z^0 \rightarrow \bar{b}b$ vertex. The 4-body phase space in soft sector possesses the form
\begin{eqnarray}
\mathrm{d}\textmd{PS}_{4}\mid_{soft}=\mathrm{d}\textmd{PS}_{3}
\frac{d^{3}p_{7}}{(2\pi)^{3}2p_{7}^{0}}\mid_{p_{7}^{0}<\delta}\ . \label{eqr:4}
\end{eqnarray}
Finally in the small $\delta$ limit, the $\mathrm{d}\Gamma^{IR-soft}_{Real}\mid_{p_{7}^{0} < \delta}$ can be expressed as
\begin{eqnarray}
\mathrm{d}\Gamma_{Real}^{IR-soft}\mid_{p_{7}^{0}<\delta}= \mathrm{d}\Gamma_{Born}\frac{C_F\alpha_s}
{\pi}\left(\frac{1}{\epsilon^{'}_{IR}}-\ln(\delta^2)\right)
\left(1-\frac{\ln{\frac{1+x_s}{1-x_s}}}
{2x_s}\right)+\mathrm{finite~terms} \label{eqr:5}
\end{eqnarray}
with $x_s=\sqrt{1-\frac{4mb^2mc^2}{(s_2-mb^2-mc^2)^2}}$. Here, those $1/\epsilon^{\prime}_{IR}$ involved terms in $(\ref{eqr:5})$ will cancel the IR singularities in CTs, i.e. $(\ref{eqv:3})$, and those remaining ones in the one-loop Feynman diagrams. Note that the $\ln(\delta^2)$ involved terms will be canceled by the $\delta$-dependent terms in the hard sector. In the end, the final result will be IR finite.

In the case of hard gluons, the decay width reads
\begin{eqnarray}
\mathrm{d}\Gamma_{Real}^{IR-hard}\mid_{{p_7}^0>\delta}= \frac{1}{2m_{Z}}\frac{1}{3}\sum|
{\cal M}^{IR-hard}_{Real}|^{2}\; \mathrm{d} \textmd{PS}_{4}\mid_{hard}\ .\label{eqr:6}
\end{eqnarray}
Here in $|{\cal M}^{IR-hard}_{Real}|^{2}$, the radiation gluon is considered to be hard, and the phase space $\mathrm{d}\textmd{PS}_{4}\mid_{hard}$ can be expressed as
\begin{eqnarray}
\int\mathrm{d}\textmd{PS}_{4}\mid_{hard}&=&\frac{2}{(4\pi)^{6}}\frac{
\sqrt{(sy-m_{c}^{2}-m_{b}^{2})^{2}-4m_{c}^{2}m_b^2}}{y} \int_{m_{B_c^*}}^{\frac{\sqrt{s}}{2}}\mathrm{d}{p_{0}}^{0}\int_{-1}^{1}
\mathrm{d}\cos\theta_{c}\int_{0}^{2\pi}\mathrm{d}\phi_{c}\nonumber\\
&&\times\left\{\int_{\delta}^{{p_{7}}^{0}_{-}}\mathrm{d}{p_{7}}^{0}
\int_{y_{-}}^{y_{+}}\mathrm{d}y+\int_{{p_{7}}^{0}_{-}}^{{p_{7}}^{0
}_{+}}\mathrm{d}{p_{7}}^{0}\int_{\frac{(m_{B_c^*})^{2}}{s}}^{y_
{+}}\mathrm{d}y\right\}\ , \label{eqr:7}
\end{eqnarray}
with
\begin{eqnarray}
&&{p_{7}}^{0}_{+}= \frac{s-2\sqrt{s}{p_{0}}^{0}}
{2(\sqrt{s}-{p_{0}}^{0} -|\overrightarrow{{p_{0}}|})}\ ,\nonumber\\
&&{p_{7}}^{0}_{-} = \frac{s-2\sqrt{s}{p_{0}}^{0}}
{2(\sqrt{s}-{p_{0}}^{0} + |\overrightarrow{{p_{0}}}|)}\ ,\nonumber\\
&&y_{+}=\frac{1}{s}[(\sqrt{s}-{p_{0}}^{0}-{p_{7}}^{0})^{2}-|
\overrightarrow{{p_{0}}}|^{2}-({p_{7}}^{0})^{2}
  +2|\overrightarrow{p_{0}}|{p_{7}}^{0}]\ , \nonumber\\
&&y_{-}=\frac{1}{s}[(\sqrt{s}-{p_{0}}^{0}-{p_{7}}^{0})^{2}-|
\overrightarrow{{p_{0}}}|^{2}-({p_{7}}^{0})^{2}
  -2|\overrightarrow{p_{0}}|{p_{7}}^{0}]\ .\label{eqr:8}
\end{eqnarray}
Here, $y$ is a dimensionless parameter defined as $y =
(k-p_0-p_7)^2/s$, $|\overrightarrow{p_{0}}|=\sqrt{({p_0}^0)^{2}-(m_{{B}^*_{c}})^{2}}$ and $\sqrt{s}=m_{Z}$.

The IR-free decay width with real correction can be formulated as
\begin{eqnarray}
\mathrm{d}\Gamma_{Real}\mid_{IR-free}=\frac{1}{2m_{Z}}\frac{1}{3}\sum|
{\cal M}^{IR-free}_{Real}|^{2}\; \mathrm{d} \textmd{PS}_{4}\ , \label{eqr:9}
\end{eqnarray}
where $|{\cal M}^{IR-free}_{Real}|^{2}$ is the amplitudes squared without IR singularities, and $\mathrm{d} \textmd{PS}_{4}$ has exactly the same form of $\mathrm{d} \textmd{PS}_{4}\mid_{hard}$ while $\delta=0$. Finally, the sum of the soft and hard sectors, i.e. (\ref{eqr:5}) and (\ref{eqr:6}), together with the IR-free part (\ref{eqr:9}) give the full real correction.

\vspace {7mm}
In the end, with real and virtual corrections, one can readily obtain the total decay width for the inclusive process $Z^{0}(k)\rightarrow {B}^*_c(p_{0})+\bar{c}(p_{5})+b(p_{6})$ at the NLO accuracy of QCD,
\begin{eqnarray}
\Gamma_{total} = \Gamma_{Born}+\Gamma_{Virtual}+\Gamma_{Real}+
{\mathcal{O}}(\alpha_s^4)\ .\label{eqtotal}
\end{eqnarray}
In (\ref{eqtotal}), the decay width $\Gamma_{total}$ is UV and IR finite, and technical cut $\delta$ independent as expected.

\section{Some Technical Details in the Calculation}

In the conventional dimensional regularization scheme, the $\gamma^{5}$ problem needs to be handled carefully, especially in the process which contains the axial-vector current. In this work, we adopt the scheme given in Ref. \cite{gamma5}, where the following rules are followed.

I. The anticommutation relations, i.e. $\{\gamma^{\mu}, \gamma^5\}=0$ and $\{\gamma^{\mu}, \gamma^{\nu}\}=2 g^{\mu \nu}$.

II. The cyclicity is forbidden in traces involving odd number of
$\gamma^{5}$. When several diagrams contribute to one process, one should
write down the amplitudes starting from the same vertex, named reading point.

III. As a special case of rule II, in the anomalous axial-vector current
situation, the reading point must be the axial-vector vertex in order to guarantee the conservation of the vector current.

\begin{figure}[t,m,u]
\centering
\includegraphics[width=15cm,height=4.5cm]{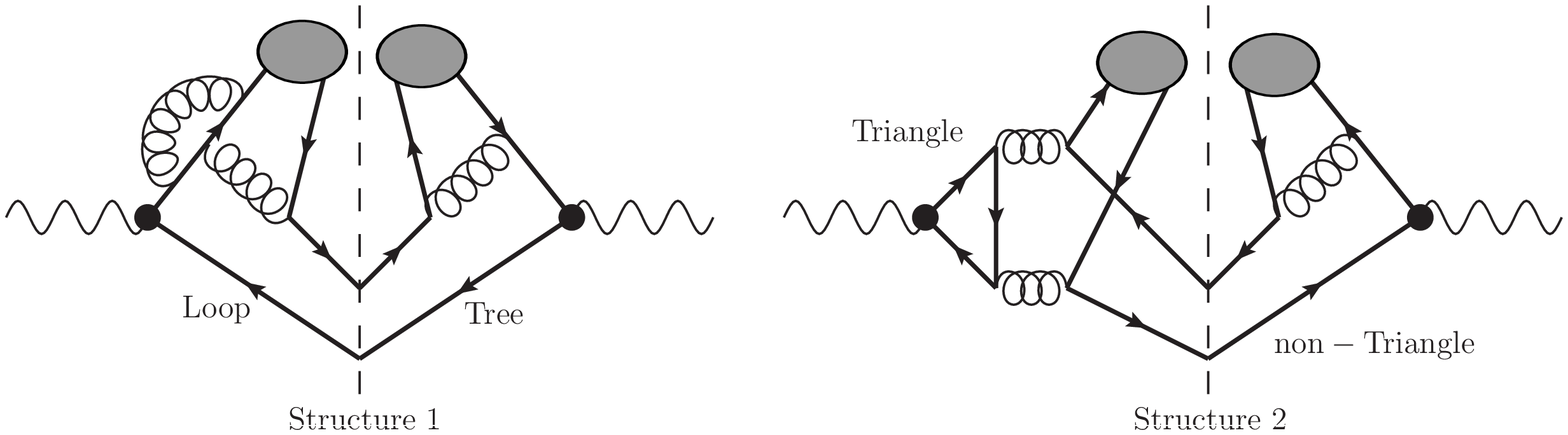}
\caption{\small Typical structures of the amplitudes squared of virtual correction.} \label{graph7}
\end{figure}

When applying these rules to our process $Z^{0}(k)\rightarrow {B}^*_c(p_{0})+\bar{c}(p_{5})+b(p_{6})$,
some conclusions are obtained. That is, in the virtual correction, the amplitudes squared have the following two structures, schematically shown in Fig.\ref{graph7}.

Structure 1: The two $Z^0-\bar{q}q$ vertexes lie in one fermion trace.

Structure 2:  There are two fermion traces, each with a $Z^0-\bar{q}q$ vertex and one involving triangle anomalous diagram, e.g.  $\mathrm{Triangle N_{11}}$ and $\mathrm{Triangle N_{12}}$ in Fig.\ref{graph4}.

For those whose amplitude squared belongs to ''Structure 1", since the amplitude can be separated into vector part and the axial-vector part,  amplitudes squared can be written as
\begin{eqnarray}
\mathrm{Trace}[{\cal M}_{Loop}{\cal M}_{Tree}^{\dag}]&=&\mathrm{Tr}[({\cal M}_{Loop}^{vector}+{\cal M}_{Loop}^{axial-vector})({\cal M}_{Tree}^{vector}+{\cal M}_{Tree}^{axial-vector})^{\dag}]\nonumber\\
&=&\mathrm{Tr}[{\cal M}_{Loop}^{vector}{{\cal M}_{Tree}^{vector}}^{\dag}]+\mathrm{Tr}[{\cal M}_{Loop}^{axial-vector}{{\cal M}_{Tree}^{axial-vector}}^{\dag}]\nonumber\\
&&+\mathrm{Tr}[{\cal M}_{Loop}^{vector}{{\cal M}_{Tree}^{axial-vector}}^{\dag}]+\mathrm{Tr}[{\cal M}_{Loop}^{axial-vector}{{\cal M}_{Tree}^{vector}}^{\dag}]\ . \label{eqd:1}
\end{eqnarray}
In the last equation, though the first trace has no $\gamma^5$ and the second trace has two $\gamma^5$s, we can move them together by anticommutation relation and contract them  to unit 1; for the other two terms each has one $\gamma^5$, and they will be canceled by their own complex conjugate terms. As a result, there is no need to evaluate the last two traces in (\ref{eqd:1}), and the calculation is hence greatly simplified.

For the triangle anomalous diagrams case, or amplitudes squared in ''Structure 2", as shown in Fig.\ref{graph7} the amplitude squared reads
\begin{eqnarray}
{\cal M}_{Anomalous}{\cal M}_{Born}^{\dag}&=&\mathrm{Trace}[{\cal M}_{Triangle}]~\mathrm{Trace}[{\cal M}_{non-Triangle}]\nonumber\\
&=&\mathrm{Tr}[{\cal M}_{Triangle}^{vector}+{\cal M}_{Triangle}^{axial-vector}]~\mathrm{Tr}[{\cal M}_{non-Triangle}^{vector}+{\cal M}_{non-Triangle}^{axial-vector}]\nonumber\\
&=&\mathrm{Tr}[{\cal M}_{Triangle}^{vector}]~\mathrm{Tr}[{\cal M}_{non-Triangle}^{vector}] \nonumber\\
&&+\mathrm{Tr}[{\cal M}_{Triangle}^{vector}]~\mathrm{Tr}[{\cal M}_{non-Triangle}^{axial-vector}] \nonumber\\
&&+\mathrm{Tr}[{\cal M}_{Triangle}^{axial-vector}]~\mathrm{Tr}[{\cal M}_{non-Triangle}^{vector}]\nonumber\\
&&+\mathrm{Tr}[{\cal M}_{Triangle}^{axial-vector}]~\mathrm{Tr}[{\cal M}_{non-Triangle}^{axial-vector}]\ . \label{eqd:2}
\end{eqnarray}
Here in the last equation, the first term will be canceled by the one whose triangle fermion loop is reversed, i.e. the vector currents of $\mathrm{Triangle N_{11}}$ and $\mathrm{Triangle N_{12}}$ cancel with each other; the second and third terms will be canceled by their complex conjugate terms; the last term survives, and it is the only one we need to trace and contains one $\gamma^5$. In the final numerical calculation, the contribution coming from triangle anomalous diagrams turns out to be numerically insignificant.

In the case of real correction, since all the amplitudes squared have the same structure as ''Structure 1", one only needs to handle the Dirac traces without $\gamma^5$ or with two $\gamma^5$s.

In our calculation, the Mathematica package FeynArts \cite{feynarts} is used to generate the Feynman diagrams,  FeynCalc \cite{feyncalc} and FeynCalcFormLink \cite{formlink} are used to handle the algebraic trace manipulation, \$Apart \cite{apart}, FIRE \cite{fire} together with codes written by ourselves are employed to reduce all the one-loop integrals into master-integrals($A_0, B_0, C_0, D_0$), and the LoopTools \cite{looptools} is employed to calculate the master-integrals numerically. The numerical
integrations of the 3- and 4-body phase spaces are performed by VEGAS \cite{vegas}.

\section{Numerical Results}

For the numerical calculation, the following input parameters are used \cite{PDG}:
\begin{eqnarray}
&&m_{c}=1.5\pm0.1~\textmd{GeV}\ ,\; m_{b}=4.9\pm0.2~\textmd{GeV}\ ,\; m_{Z}=91.1876~\textmd{GeV}\ ,\;m_{W}=80.399~\textmd{GeV}\ ,\;\nonumber\\
&&\mathrm{sin}^2\theta_W=0.2312\ ,~g=2\sqrt{2}m_W \sqrt{\frac{G_F}{\sqrt{2}}}=0.6531\ ,~\Psi_{{B}^*_c}(0)=\frac{R(0)}
{\sqrt{4\pi}}=0.3615\;\textmd{GeV}^{\frac{3}{2}}\ .\label{eqn:1}
\end{eqnarray}
Here, $G_F$ is the Fermi constant in weak interaction and $R(0)$ is the $B^*_{c}$ meson's radial wave function at the origin, which value is estimated via potential model \cite{potential}. The two-loop strong coupling of
\begin{eqnarray}
\frac{\alpha_s(\mu)}{4\pi}=\frac{1}{\beta_0 L}-\frac{\beta_1\ln
L}{\beta_0^3L^2}\
\end{eqnarray}
is employed in the calculation. In which $L=\ln(\mu^2/\Lambda_{QCD}^2)$ with $\Lambda_{QCD}$ to be $214$ MeV, and $\beta_1=(34/3)C_A^2-4C_FT_Fn_f-(20/3) C_AT_Fn_f$ is the two-loop coefficient of the QCD beta function.

\begin{figure}[t,m,u]
\centering
\includegraphics[width=8cm,height=6cm]{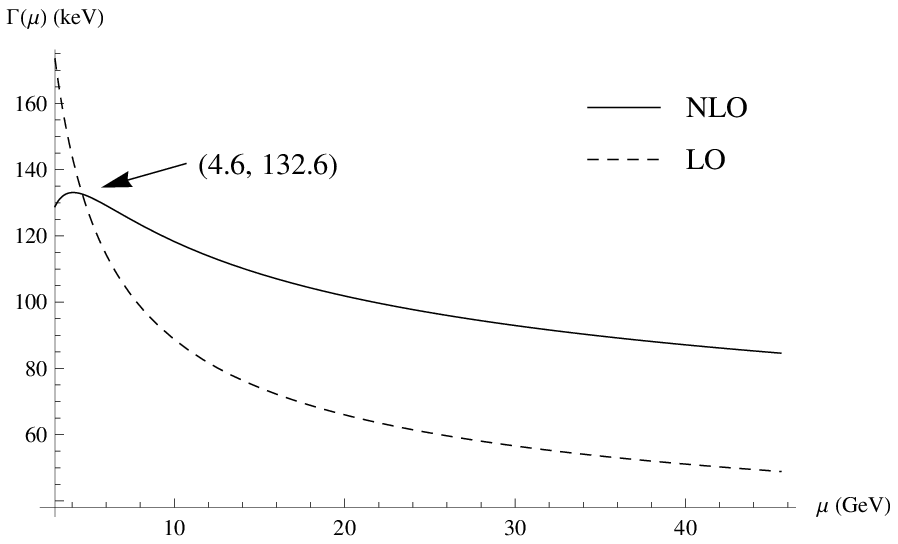}
\includegraphics[width=8cm,height=6cm]{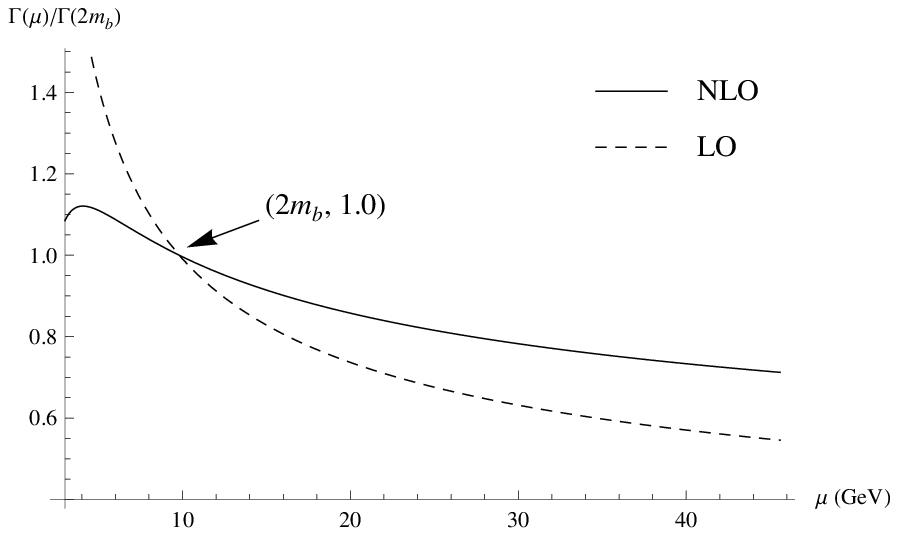}
\caption{\small The decay width $\Gamma(\mu)$ and the ratio $\Gamma(\mu)/\Gamma(2m_b)$ versus renormalization scale $\mu$ for $\mu$ running from $2m_c$ to $m_Z/2$.} \label{graph8}
\end{figure}

In Fig.\ref{graph8}, the decay width $\Gamma(\mu)$ and the ratio $\Gamma(\mu)/\Gamma(2m_b)$ versus renormalization scale $\mu$ are presented. The NLO corrections enhance the LO contribution to $Z^{0}\rightarrow {B}^*_c+\bar{c}+b$ when $\mu > 4.6$ GeV. We can see from  Fig.\ref{graph8} that the renormalization scale dependence of the decay width is reduced evidently.
\begin{table}[h]
\caption{Decay widths $\Gamma(\mu)$(keV) of
$Z^{0}\rightarrow {B}^*_c+\bar{c}+b$ at leading
order and next-to-leading order. The first error are caused by
$m_c=1.5\pm 0.1$ GeV and the second ones by $m_b=4.9\pm 0.2$ GeV at $\mu=2 m_b$, 28 GeV, $m_Z/2$.}
\begin{center}
\renewcommand\arraystretch{1.4}
  \begin{tabular}{|l||c|c||c|}
    \toprule
    $\ \ \Gamma(\mu)$(keV)& LO & NLO & K-factor\\
    \hline
    $\mu = 2m_b$& $89.56^{-17.10+0.71}_{+22.85-0.71}$ & $118.77^{-23.11+0.37}_{+31.53+0.07}$ & $1.33^{-0.01-0.01}_{+0.01+0.01}$ \\
    $\mu = 28$& $58.02^{-11.08+0.46}_{+14.80-0.46}$ & $94.40^{-18.25+0.45}_{+24.73-0.22}$ & $1.63^{-0.00-0.01}_{+0.01+0.01}$ \\
    $\mu = m_Z/2$& $48.87^{-9.33+0.39}_{+12.47-0.39}$ & $84.60^{-16.33+0.44}_{+22.08-0.26}$ & $1.73^{-0.00-0.00}_{+0.01+0.01}$\\
    \botrule
  \end{tabular}
\end{center}
\label{tab:1}
\end{table}

We adopt three typical renormalization scales in evaluation, and the corresponding values of running coupling constant are $\alpha_s(\mu=2m_b)=0.1768$, $\alpha_s(\mu=28$ GeV) = 0.1423 and $\alpha_s(\mu=m_Z/2)=0.1306$, the decay widths are as presented in Tab.\ref{tab:1}, where the first errors are induced by $m_c=1.5 \pm 0.1$ GeV and the second ones are induced by $m_b = 4.9 \pm 0.2$ GeV. Here, the K-factor is defined as $\frac{\Gamma_{NLO}}{\Gamma_{LO}}$. In the calculation, when estimate the uncertainties induce by varying the $m_c$, the $m_b$ is fixed and vice verse.

\begin{figure}[t,m,u]
\centering
\includegraphics[width=8cm,height=6cm]{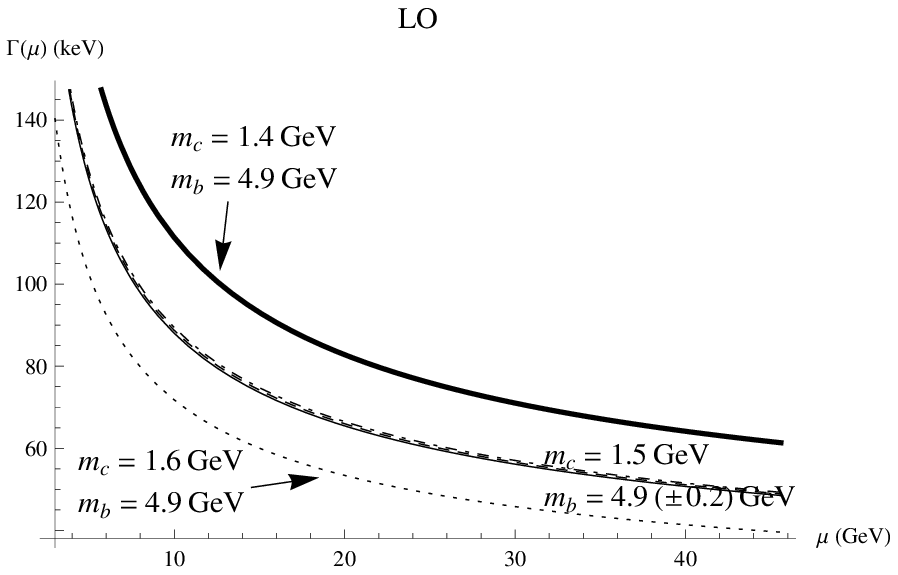}
\includegraphics[width=8cm,height=6cm]{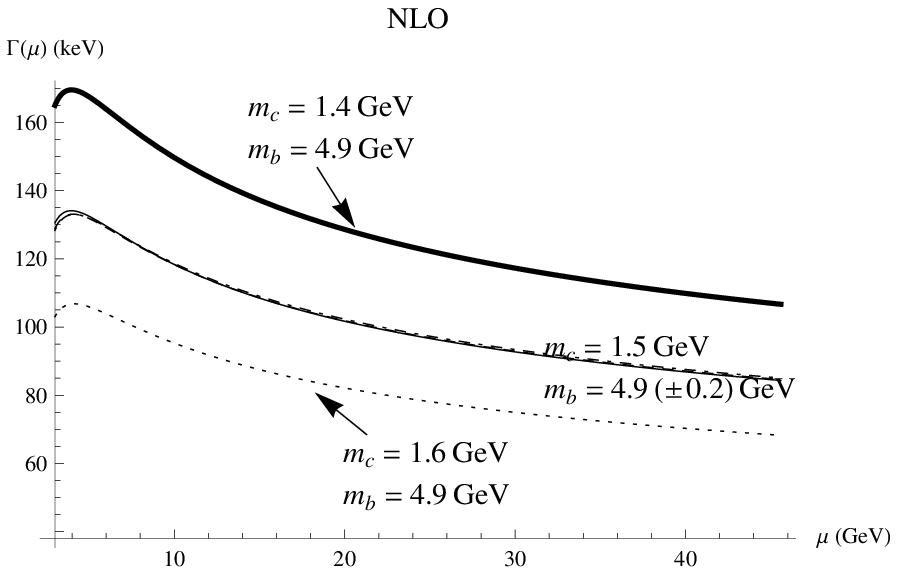}
\caption{\small The decay widths of LO(left) and NLO(right) with different  quark masses versus running renormalization scale $\mu$, from $2m_c$ to $m_Z/2$.} \label{graph9}
\end{figure}

The results of Table \ref{tab:1} indicate that
\begin{itemize}
\item The correction of NLO is significant, and the K-factor grows with the renormalization scale $\mu$ increase, yet it grows slower at high scale region;
\item When renormlization scale $\mu$ increases, the decay widths for both LO and NLO decrease, yet both decrease slower at high scale region;
\item The uncertainties caused by varying $m_c$ are much larger than those induced by $m_b$.
\end{itemize}
To show the above results more clearly, in Fig.\ref{graph9} we exhibit the LO(left) and NLO(right) decay widths with different quark masses versus running renormalization scale $\mu$. It is obvious that the decay widths are much more sensitive to $m_c$ then $m_b$, the three lines of $m_b=4.9(\pm0.2)$ are very close to each other at both LO and NLO.

According to Ref. \cite{slp} and above calculation, the $B_c$ and $B^{(*)}_c$ production rates in $Z^0$ decays are readily obtained in LO and NLO, at scale $\mu=2m_b$ for illustration. That is
\begin{eqnarray}
Br_{LO}(Z^0 \to B_c+\bar{c}b)&=&\frac{\Gamma_{LO}(B_c)}{\Gamma_Z}=2.9\times10^{-5}\ ,\nonumber\\
Br_{NLO}(Z^0 \to B_c+\bar{c}b)&=&\frac{\Gamma_{NLO}(B_c)}{\Gamma_Z}=3.1\times10^{-5}\ ,\nonumber\\
Br_{LO}(Z^0 \to B^*_c+\bar{c}b)&=&\frac{\Gamma_{LO}(B^*_c)}{\Gamma_Z}=3.6\times10^{-5}\ ,\nonumber\\
Br_{NLO}(Z^0 \to B^*_c+\bar{c}b)&=&\frac{\Gamma_{NLO}(B^*_c)}{\Gamma_Z}=4.8\times10^{-5}\ ,\label{eqn:2}
\end{eqnarray}
where $\Gamma_Z=2.5~\textmd{GeV}$ is the total decay width of $Z^0$ boson. Because the spin-triplet state $B^*_c$ decays to the ground state $B_c$ with  almost 100\% rate, we obtain the total production rates for $B_c$ production in $Z^0$ decay,
\begin{eqnarray}
Br_{LO}(Z^0 \to B_c+X)&=&6.5\times10^{-5}\ ,\nonumber\\
Br_{NLO}(Z^0 \to B_c+X)&=&7.9\times10^{-5}\ .\label{eqn:3}
\end{eqnarray}
Given about $10^{9}$ or more $Z^{0}$ events being produced per year
in future colliders, there would be sizable $B_c$ being produced.
And the effects of the NLO corrections might be captured in
experimental observation. It is worth noting that the LO
contributions from P-wave/color-octet are an order of magnitude
smaller than the S-wave \cite{Z2}, which only enhance the total
production rates slightly.

\section{Summary and Conclusions}

In this work we calculated the decay width of $Z^{0}\rightarrow {B}^*_c+\bar{c}+b$ inclusive process at the NLO accuracy in the framework of NRQCD. The calculation procedures for both LO and NLO corrections were presented. Our analyses for the vector and axial-vector currents contributions were performed in dimensional regularization scheme. The decay width and its uncertainties caused by varying quark masses, as well as the dependence on renormalization scale $\mu$ were presented numerically. Supposing that there will be copious $Z^{0}$ produced in future colliders, our calculation together with previous work \cite{slp} would be helpful for the precise study of $B_{c}$ physics, and might also tell how well the perturbative calculation and non-relativistic quark model
work for $B_{c}$ system.

According to the calculation, the NLO QCD correction to the inclusive process $Z^{0}\rightarrow {B}^*_c+\bar{c}+b$ is significant. We found that the renormalization scale $\mu$ dependence of the decay width is depressed while the NLO correction is taken into account.
When the scale $\mu$ runs from $2m_b$ to $m_Z/2$, both values of $\Gamma_{LO}$ and $\Gamma_{NLO}$ decrease, yet the ratio $\Gamma_{NLO}/\Gamma_{LO}$ grows from $1.33$ to $1.73$, and the increasement trend slows down at high $\mu$ region. Moreover, the input parameter of quark mass $m_c$ has a quite large influence on the decay width, for both $\Gamma_{LO}$ and $\Gamma_{NLO}$, at both low and high $\mu$ regions as well. In contrast to $m_c$, the uncertainties induced by $m_b$ are negligible.

\vspace{.7cm} {\bf Acknowledgments}

This work was supported in part by National Key Basic Research Program of China under the grant 2015CB856700, and by the National Natural Science Foundation of China(NSFC) under the grants 11175249, 11121092, and 11375200.


\end{document}